\documentstyle[11pt,paspconf,epsf]{article}
\makeindex

\begin{document}


\def\myfig #1#2#3#4{\par
\epsfxsize=#1 cm
\moveright #2cm
\vbox{\epsfbox{#3}}
{\noindent Figure~#4 }\vskip .3cm }
\def\bh{black hole}
\def\ew{equivalent width~}
\def\ad{accretion~disk~}
\def\el {emission-line}
\def\bhs{black~holes}
\def\ar{accretion~rate}
\def\ip{ionization-parameter~} 
\def\ed {Eddington~}
\def\ers{{\rm erg/sec}}
\def\ms{M_{\odot}}
\def\et{{\it et.al.}}
\def\sw{Schwartzshild~}
\def\bb{black body~}
\def\be{Baldwin effect~}

\title{Evolutionary Baldwin Effect in AGN
}


\author{A. Wandel}
\affil{ Racah Institute, The Hebrew University, Jerusalem 91904, 
Israel}

\begin{abstract}

Assuming Active Galactic Nuclei are powered by accretion onto a massive \bh~
we suggest that
the growth of the central \bh~ mass due to the matter accreted over
the AGN  lifetime causes evolution of the luminosity and spectrum.

We show that the effective temperature of the UV continuum spectrum is
likely to be anti-correlated with the \bh~ mass and with the luminosity.
We estimate the change in the \ew of the emission lines due to the
growth of the \bh~ and
show that for plausible evolutionary tracks and effective temperature models
the \ew is anti-correlated with continuum luminosity
thus implying an evolutionary origin to the Baldwin effect.

\end{abstract}


\section{Introduction}

The most accepted explanation for the energy source of quasars and Active
Galactic Nuclei (AGN) is the accretion of matter onto supermassive \bh s
in the center of the host galaxy.
The amount of accreted matter required to explain the the observed energy
output is 0.001-few solar masses per year, while the
central mass required to maintain steady accretion (the
Eddington limit) is of the order of $10^6-10^9 \ms {\rm yr}^{-1}$.
Comparing these two numbers gives an age of $\sim 10^7 - 10^9$ years, which
indicates a  timescale for the evolution of the \bh~ mass and
 the AGN properties.
  
Conservation of angular momentum directs
the accreted material to form a disk around the central \bh~
and the properties of such an accretion disk may be calculated under a few
basic assumptions (e.g. Sakura and Sunyaev, 1973). In particular one can
calculate the luminosity and spectral distribution of the radiation from
the accretion disk, given the \bh~ mass and the \ar .

As the \bh~ mass grows due to the accreted material, the  luminosity and
the emerging spectrum  change with time. 
Since the continuum radiation ionizes the line-emitting material and
drives the broad AGN \el s, the change in the properties of the
radiation from the central source changes also the properties
of the emission lines. In particular the spectral shape of the ionizing 
radiation would determine the \ew   (the ratio between
the energy flux in the line to the continuum energy flux at the line 
frequency) of the \el s.

Baldwin (1977) showed that the \ew of the CIV \el~ decreases with increasing 
continuum luminosity, a correlation known as the 'Baldwin effect'. A similar
relation has been found also for other broad \el s. The origin of this effect 
is not understood.
 
In this work we suggest that at least part of the \be can be attributed to 
evolution. We estimate the relation between the \ew and the continuum spectrum, and
calculating the evolution of the continuum emission from the \ad we
show that for plausible evolutionary tracks the \ew is anti-correlated
with the continuum luminosity.

\section{Basic relations for accreting \bh s}

The basic properties of a compact, accretion powered radiation source must obey several basic 
relations:

\smallskip
{\bf a. The \ed limit:} 
in order to maintain steady spherical accretion the luminosity must be less than 
the \ed luminosity, 
$$L<L_{Edd}=4\pi GMm_pc/\sigma_T=1.3 10^{46}M_8 ~\ers
$$
or
\begin{equation}
\label {equ:med}
 M_8=0.7 \eta^{-1} L_{46}
\end{equation}

where $ M_8=M/10^8\ms$, $\eta=L/L_{Edd}$ is the \ed ratio, and $L_{46} =L/10^{46}$ \ers .

\smallskip
{\bf b. The \bb temperature:} 
if a luminosity $L$ comes from a region of radius R and 
temperature $T$, then  
\begin{equation}
\label {equ:lrt}
L<4\pi R^2\sigma T^4.
\end{equation}
 If a spectral feature at a photon
energy of $E$ is due
to \bb emission from a region of size $R$, then the temperature is given by 
$E\approx 3kT$ and reversing the relation above
we have an upper limit on the \bh~ mass (for a true \bb spectrum this
becomes an approximate equality):

\begin{equation}
\label {equ:mt}
M_8 > 130 E_{Ryd}^{-2}{L_{46}}^{1/2} (R/R_s)^{-1}
\end{equation}

where $R_s=2GM/c^2\approx 3~10^{13} M_8~$cm is the \sw radius
and $E_{Ryd}$ is the spectral energy in Rydbergs.

Combining eqs. \ref{equ:med} and \ref{equ:mt} we can eliminate $M$
obtaining (for \bb emission) an estimate of the size:

\begin{equation}
\label {equ:rle}
{R\over R_s}\approx 200 L_{46}^{-1/2} \eta {E_{Ryd}}^{-2}
\end{equation}

\smallskip
{\bf c. Variability:}
the shortest time scale for global variations in the luminosity is the light 
travel time across the \sw radius, hence if the 
luminosity is observed to vary significantly on a time scale $\delta t$, the 
\bh~ mass has to be
\begin{equation}
\label {equ:mdt}
M_8<(\delta t/10^3 r^{-1} {\rm sec}),
\end{equation}
where $r$ is the effective radius of emission in units of $R_s$.

However, for continuum changes near the feature  of spectral energy $E$ 
the light travel time is (from eq. \ref{equ:rle})
\begin{equation}
\label {equ:trc}
\delta t_l \approx 1.3~10^5 L_{46}^{1/2}  {E_{Ryd}}^{-2} {\rm sec},
\end{equation}
and the dynamical time is 
\begin{equation}
\label {equ:trd}
\delta t_d \approx 20 L_{46}^{1/4} \eta^{1/2} {E_{Ryd}}^{-3} {\rm days},
\end{equation}
which gives time scales of the order of the observed  UV variability in 
Seyferts and quasars.
\smallskip

{\bf d. Evolution of the \bh~ mass:}
 We may define the
\ed time $t_E$ as the time required for an accretion-powered 
object radiating at the \ed luminosity to double its mass, 
due to accreted matter,
$
t_E=Mc^2/L_E=4\times 10^8 {\rm y}.
$
The observed luminosity implies an accretion rate of
$
\dot M = 0.16 \epsilon^{-1} L_{46} \ms {\rm y}^{-1}
$
where  $\epsilon$ is the efficiency.
Combining these two expressions gives the  mass, 
accumulated if the accretion rate is maintained during a time $t_E$ :
\begin{equation}
\label {equ:mev}
M_{8} = 0.6 \epsilon^{-1} (L/L_{Edd})^{-1} f_L L_{46} 
\end{equation}
where  $f_L$  is the fraction of the active 
time for intermittently active AGN.

\section{The thin accretion-disk model}
Many authors tried to fit \ad spectra to the observed AGN continuum, thus finding the 
\ad parameters ($M, \dot M, \alpha$) and the angular momentum of the \bh~ (\sw or 
Kerr) which best fit 
the observed continuum in the UV  (Wandel and Petrosian 1988; Sun and Malkan 
1991) or in the soft X-rays (Laor  1990). 
\subsection{The \ad spectrum}

The radiative energy output of the \ad is dominated by the release of gravitational 
potential energy, the rate of which is $GM\dot M/r$. A more accurate calculation yields
\begin{equation}
\label {equ:lr}
L(R) = {3GM\dot M\over 8\pi R^3} \left [ 1-\left ({R_{in}\over R }\right )^{1/2}\right ]
\end{equation}

where $R_{in}$ is the inner disk radius.
In the outer part of the thin disk the opacity is dominated by true absorption
and the local spectrum is a \bb spectrum. For this part of the disk,
comparing eq. \ref{equ:lr} to the \bb radiation flux gives for the disk surface temperature
\begin{equation}
\label {equ:tr}
T(R) \approx \left (  {3GM\dot M\over 8\pi\sigma R^3} \right )^{1/4} 
\approx 6 ~10^5 \left (  {\dot m\over M_8} \right ) ^{1/4} r^{-3/4}~{\rm K}
\end{equation}
\noindent
where $r=R/R_s$ and  $\dot m=\dot M/\dot M_{Edd}$.
When the \ar~ approaches the \ed rate, the intermediate disk region becomes 
dominated by electron-scattering, the spectral function will be of a modified \bb
and the surface temperature is higher than given by eq. \ref{equ:tr}.
At still smaller radii, the pressure in the disk is dominated by radiation, rather than by 
gas pressure, and the thin disk solution becomes thermally unstable.
In that inner region the thin disk solution is probably not valid, and has to be replaced
by a hot disk solution (e.g. Wandel and Liang 1991).
It turns out that the intermediate modified \bb region is relatively narrow, 
and close to the \bb - electron scattering boundary the spectrum is nearly a \bb one,
so both regimes may be approximated by the \bb solution.

The spectrum of the \bb (and modified \bb ) disk is given by integrating over the entire
disk, 
\begin{equation}
\label {equ:fn}
F(\nu ) \approx \int_{R_t}^{R_{out}} 2\pi R B_\nu[T(R) ] dR
\end{equation}

where $B_\nu(T)$ is the Planck function and
$R_t$ is the transition radius from the intermediate to the inner radiation-pressure
dominated region (for $\dot m<0.02$ the \bb region extends down to the inner edge of the disk
and $R_t=R_{in}$ ).

Since the $B_\nu (T)$ has a sharp peak at $h\nu_{max}\approx 3kT$ and cuts off at higher 
frequencies, the highest frequency of the \bb part of the disk spectrum comes from the 
radius $R_t$, with the highest temperature for which the disk is still optically thick. 
Eq. \ref{equ:fn}
gives a spectrum which depends on the radial extent of the \bb part of the disk.
If that part is extended ($R_{out}/R_t>>1$) the spectrum is almost flat ($F(\nu )\sim \nu^{1/3}$)
and cuts off beyond $\nu_{max}\approx 3kT(R_t)/h$ (or $3kT(5R_S)/h$ for $\dot m <0.1$).
If $R_{out}/R_t\sim$ a few,  the spectrum
will be merely a somewhat broadened  Planck spectrum.

While in the UV band the thin \ad multiple \bb spectrum may be a good 
approximation, the soft X-rays may 
be produced by a hotter medium due to processes other than \bb emission, such as 
Comptonization, a two-temperature disk (Wandel and Liang 1991) or hot corona 
(Haardt, Maraschi and Ghisellilni 1994; Czerny, Witt and Zycki 1996).

\section{Deriving $M$ and $\dot M$ from the UV spectrum}

We want to find an approximate relation between the \bh~ mass and \ar~ and the 
continuum spectrum.
The \ar~ is determined straightforwardly by the luminosity, via the relation
$L_{ion}=\epsilon f_{ion} \dot M c^2$, where $f_{ion}$ is the bolometric correction
for the ionizing continuum.
To estimate the \bh~ mass we may use the multi-\bb~\ad -spectrum.
As discussed in the previous subsection  the UV bump cutoff frequency is determined by the 
highest surface temperature in the \bb part of the disk.
If the \bb region extends down to a radius $R_t$, and the EUV cutoff energy is $E_{co}$,
then  eq. \ref{equ:mt} gives (we assume the inner region is optically thin and much hotter, 
so its contribution in the frequencies of interest can be neglected)
\begin{equation}
\label {equ:mte}
M_8\approx 1 (E_{co} /3 eV )^{-2}L_{46}^{1/2} (R_t/R_s)^{-1}.
\end{equation}

In several models the \bb regime extends close to the inner disk edge.
This is the case in the ``$\beta$'' disk model with viscosity proportional to the gas
pressure ($\nu\sim\beta P_{gas}$). For low \ar s ($\dot m<0.1$) this is true also for the
``$\alpha$'' model with viscosity proportional to the total pressure
($\nu\sim\alpha (P_{gas}+P_{rad})$)
In this case, since the disk emissivity peaks at $R\approx 5R_S$ 
(for a non-rotating \bh ) we may
use eq. \ref{equ:mte} with $R/R_s=5$ in order to find the \bh~ mass:
$M_8\approx 3 L_{45}^{1/2}/ (E_{co} /3 eV )^2 $ .

It is possible to do a more refined treatment, calculating for each pair of \ad 
parameters $M$ and $\dot M$ fitting not the total luminosity and blue-bump 
temperature, but the actual observables, e.g. the UV luminosity and spectral index. 
Wandel and Petrosian (1988) have calculated the \ad flux and spectral index 
 at 1450\AA , for a grid of \ad parameters $(M, \dot M)$. Inverting the grid they have obtained 
contours of constant \bh~ mass 
 and constant \ed ratio ($\dot m$) in the $L-\alpha_{UV}$ plane (fig 1) .
 Plotting in this plane samples of AGN it is  possible to read off the contours the 
corresponding \ad parameters. Comparing several groups of AGN a systematic trend 
appears: higher redshift and more luminous objects tend to have larger \bh~ masses 
and luminosities closer to the \ed limit (see table~1). 
 Similar results are obtained by Sun and Malkan (1991).
\begin{table}
\caption{Grouping of AGN \ad parameters.} \label{tbl1}
\begin{center}
\begin{tabular}{lccc}
\tableline
{}&{}&{}&{}\\
AGN group & Log$ F_{1450}$ & Log $M/\ms$ & $\dot M/\dot M_{Edd}$ \\ 
{}&{}&{}&{}\\
\tableline
{}&{}&{}&{}\\
Seyfert galaxies & 28-29.5 &7.5-8.5 & 0.01-0.5\\
Low Z quasars & 29-30.5 &8-9 & 0.02-0.1\\
Medium Z quasars & 30-31.5 & 8-9.5 & 0.1-0.5\\
High Z quasars & 31-32 & 9-9.5 & 0.03-2\\
{}&{}&{}&{}\\
\tableline
\end{tabular}
\end{center}
\end{table}

\myfig {10} 1 {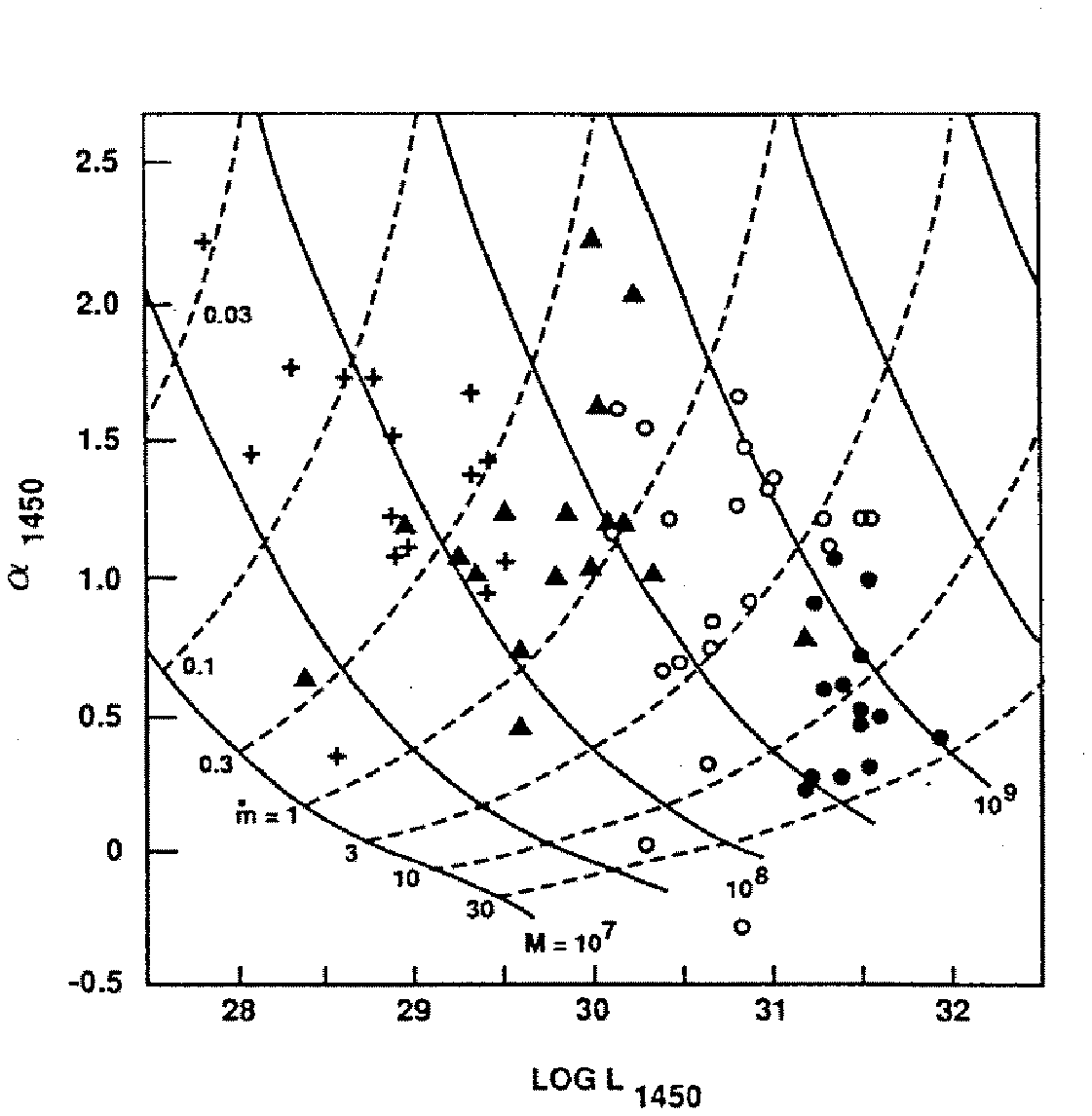}
{1. Accretion-disk \bh~ evolution tracks in the $\alpha -L$ plane (Wandel and Petrosian 1988). 
Crosses: Seyfert galaxies, triangles, open circles and filled circles:
low, medium and high redshift quasars, respectively. Continuous lines: constant mass;
dashed lines: constant $\dot M/M$ (note that  the labeling
of the curves in the figure follows the notation 
$\dot m=16.7\dot M/\dot M_{Edd}$).}

\section{Relating the spectrum to the \ew }

The \ew of an \el~ is actually the ratio between the luminosity in the line and the continuum
luminosity at the \el~ frequency. 
The line luminosity is determined by the photoionizing
flux, near 1Ryd (13.6 eV). 
Since  $E_{max}\approx 3kT_{eff}$ is just below 13.6 eV 
(see below), this frequency is on the Wien part, and the flux is sensitively
influenced by the effective disk temperature, $T_{eff}$.
In the parameter range of interest we may therefore assume that the \ew is strongly correlated 
with $T_{eff}$. Quantitatively, the correlation is related to the logarithmic derivative
\begin{equation}
\label {equ:lblt}
{d~ln~B_\nu\over d~ln~T}= {-xe^x\over e^x-1}
\end{equation}
where $x=h\nu /kT$. For $kt<$1 Ryd, which is true in a wide range of cases, as we show below, 
eq. \ref{equ:lblt} gives 
$${d~ln~B_\nu\over d~ln~T}\approx -h\nu/kT,$$ 
namely a strong correlation with T.

In order to relate the UV ionizing spectrum near 1Ryd to the accretion parameters,
it is necessary to determine the effective \bb temperature of the emitting region.
\subsection{A general Effective Temperature Estimate}

We have already derived a quite general temperature-radius relation for
\bb emission from an accretion flow (eq. \ref{equ:rle}).
We can apply this relation to determine the temperature or radiation peak
energy as a function of the observed luminosity and estimated mass.

In order to determine a characteristic temperature we need either indicate
the size of the emitting region (and use eq. \ref{equ:lrt})
or assume an emission model, as done below for the \ad model.
The size of the emitting region can be bounded by variability analyses.
From eqs. \ref{equ:lrt} and \ref{equ:mdt} we have
\begin{equation}
\label {equ:elt}
E_{max}> 10 {L_{46}}^{1/4} (t_{day})^{-1/2} eV
\end{equation}
If we assume a linear relation between luminosity and variability time,
for example, that for objects with a continuum luminosity $L\sim 10^{46}\ers$ 
the  UV variability time is one month, eq. \ref{equ:elt} gives 
\begin{equation}
\label {equ:el}
E>2 {L_{46}}^{-1/4} eV.
\end{equation}

\subsection{Effective Temperature Models for an \ad}
In the framework of the thin \ad model, 
we consider four models for determining the effective temperature.

{\bf a. The inner disk \bb temperature}

When the disk is nearly \bb up to the inner edge, 
we can approximate the spectrum  by \bb emission near the 
maximum-emissivity radius, which gives
\begin{equation}
\label {equ:ebb}
E_{max}\approx 3kT_{BB}(5R_s)=10 (\dot m /M_8 )^{1/4}  eV.
\end{equation}
 
As discussed in sec. 3 above, this model applies to the ``$\beta$'' disk and to low
\ar s ($\dot m<0.1$) in the $\alpha$ disk.

{\bf b. The inner disk radiation-pressure temperature}

This model applies to high \ar s ($\dot m>0.1$) in the $\alpha$ disk, provided
the inner disk maintains the temperature profile given by the thin-disk solution.
The temperature in the inner, radiation dominated region is given by
(Sakura and Sunyaev 1973)
\begin{equation}
\label {equ:trad}
T_{rad}(r)\approx 2~10^5 (\alpha M_8 )^{-1/4} r^{-3/8} [1-(6/r)^{1/2}]^{1/4}~K
\end{equation}

Substituting the maximal-emissivity radius $r=5$ gives

\begin{equation}
\label {equ:erad}
E_{max}\approx 4 (\alpha M_8 )^{-1/4}  eV
\end{equation}
The gas-radiation pressure boundary -
 taking as $T_{max}$ the temperature at the inner boundary between the \bb and 
the radiation-dominated disk 

{\bf c. The \bb-radiation boundary temperature}

The effective temperature is taken at the boundary between the intermediate
and inner disk regions, at $r_{mi}\approx 10(\alpha M_8)^{0.1}\dot m^{0.76}$.

The energy of the peak in the spectrum is then derived by substituting $r_{mi}$ 
into eq. \ref{equ:trad}, which gives
\begin{equation}
\label {equ:eradb} 
E_{max}\approx 2 (\alpha\dot m M_8)^{-0.3}~ eV.
\end{equation}

This model is adequate in the case of a high \ar~ - $\alpha$ disk when the inner disk
does not follow the thin disk model but establishes a stable hot solution
(e.g. Wandel and Liang 1991).

 {\bf d. The ``photospheric'' temperature}

In this model we consider as the disk effective temperature 
the temperature at the radius where the disk 
becomes effectively optically thin.
The effective optical depth in the disk is given by

$$\tau_*=h\rho (\kappa_{es}\kappa_a)^{1/2},$$
where $h$ and $\rho$ are the disk vertical scale-height and density, respectively,
and $\kappa_{es}$ and $\kappa_a$ are the electron-scattering and absorption opacities.
Equating $\tau_*$ to unity and eliminating $T$ and $r$ by using the temperature profile 
$T(r)$ gives
\begin{equation}
\label {equ:eradph} 
E_{max}\approx 2 \alpha ^{-0.4}(\dot m M_8)^{-0.26}~ eV.
\end{equation}

Note that the last two models give very similar results.

\section {Evolutionary Scenarios}
The evolution of the \bh~ mass over cosmological times depends on the \ar , which may of
course change over time. From the \ad efficiency and the observed luminosity we can estimate 
the present \ar . In order to determine the\bh~ evolution, we need to specify the variation
of the \ar~ over time. We consider three representative scenarios:
\begin {itemize}
\item 
constant \ar
\item 
\ed limited accretion (constant \ed ratio)
\item 
spherical accretion from a homogeneous medium 
\end {itemize}

\subsection {Constant \ar}
This would correspond to external feeding of the \bh , 
which is regulated by an external potential, e.g. by stellar encounters, 
where the stellar motions are governed by the gravitational potential in the bulge
of the host galaxy, far from the  influence of the \bh .
As the mass of the \bh~ grows, $\dot M$ does not change, and hence
the accretion parameter, $\dot m\sim\dot M/M\sim M^{-1}$.
Eventually \ar~ will become very sub-\ed , and the \bh~ will be starved,
or at least on a diet.
Since $L\sim \dot M$, in this scenario the (bolometric) luminosity also is also constant.
Note that the spectral energy distribution does change (though slowly) as $M$ increases.

\subsection {\ed limited accretion}
This would be the case in a \bh~ that is over-fed.
The \ar~ cannot become very super-\ed , and the source will regulate the \ar~ to 
$L\approx L_{Edd}$ or
$\dot m\approx \epsilon^{-1}$.
Since $\dot m\sim \dot M/M$, $\dot m \sim$ constant implies
$\dot M\sim M$ and hence
$L\sim M$. 

\subsection {Spherical accretion from a homogeneous medium}
In this case we assume the matter supply comes from a homogeneous distribution 
(gas or stars), due to the gravitational potential of the \bh .
On large scales, the accretion will be spherical, and the accretion radius at which the \bh~
gravitational potential becomes significant is given by
\begin{equation}
\label {equ:rac}
R_{acc}\approx GMv_*^2 \approx 3 M_8v_{300}^2~ pc ,
\end{equation}
where
$v_*=300v_{300}$km/s is the stellar velocity dispersion.

The \ar~ is given by the Bondi formula
\begin{equation}
\label {equ:sar}
\dot M\approx 4\pi R_{acc}^2 {v_*}^2 \rho_* = (0.3 \ms /{\rm yr}) 
M_8^2v_{300}^{-3} \left ( {\rho_* \over 10\ms {\rm pc}^{-3}} \right ) ,
\end{equation}
where $\rho_*$ is the stellar number density.

In this case $L\sim\dot M\sim M^2$
and $\dot m\sim M$, that is, the \ed ratio increases with time.
Eventually the \ed ratio will approach unity and the accretion will become \ed limited.

\subsection {Accretion from an inhomogeneous medium}
A similar expression can be derived for a non-homogeneous medium, with a density profile
$\rho_*\sim R^{-p}$.
Assuming the velocity dispersion is independent of $R$,
eqs. \ref{equ:rac} and \ref{equ:sar} give for the \ar~ in that case
$$\dot M\sim M^{2-p}.$$
If also the velocity dispersion has a radial dependence,  $v_*\sim R^{-q}$
then we get a more complicated dependence, 
\begin{equation}
\label {equ:sarr}
\dot M\sim M^{{2\over 1+2q}-q-p}.
\end{equation}

In this case, any functional dependence is possible.
If for example $q=1/4$ and $ p=1/2$, 
$\dot M\sim M^{0.6}$,
and if  $q=1/2$ (point mass) and $ p=1/2$,
$\dot M =$const.

\section{Putting it all together: an evolutionary \be ?}

In the previous section we have found the relation between the accretion parameters 
$M$ and $\dot M$, and found the functional dependence of the continuum luminosity on the
evolving \bh~ mass for several accretion scenarios.
This enables us to draw evolutionary curves in the $M-\dot m$ plane (fig. 1).
For example, an \ed bound \ad will evolve along the constant $\dot m$ curves
(dashed curves in fig. 1). 
A homogeneously accreting \bh~ will evolve along curves of $\dot m\sim M$, which correspond
to nearly horizontal (from left to right) lines in fig. 1, and 
a constant \ar~ will yield $\dot m\sim M^{-1}$ which are almost vertical (upward) lines
in fig. 1.

\begin{table}
\caption{The relation between effective \ad temperature and \bh~ mass for various 
accretion scenarios} \label{tbl2}
\begin{center}
\begin{tabular}{llll}
\tableline
{}&{}&{}&{}\\
Accretion scenario: &Constant $\dot M$& Constant $L/L_{Edd}$ & Spherical \\
$L(M)\propto\dot M$&const&$\sim M$&$\sim M^2$\\
\\
\tableline
$T_{eff}$ model:&{}&{}&{}\\
{}&{}&{}&{}\\
$T_{BB}$&$M^{-1/2}$ &$M^{-1/4}$&const\\
{}&{}&{}&{}\\
$T_{rad}$&$M^{-1/4}$ &$M^{-1/4}$&$M^{-1/4}$\\
{}&{}&{}&{}\\
$T(\tau_*=1)$ or $T$(BB/rad)&const &$M^{-1/4}$&$M^{-1/2}$\\
{}&{}&{}&{}\\

\tableline
{}&{}&{}&{}\\
$T$(BB-var)$\propto L^{-1/4}$&const &$M^{-1/4}$&$M^{-1/2}$\\
{}&{}&{}&{}\\
\tableline
\end{tabular}
\end{center}
\end{table}

Combining this with the dependence of the effective radiation-temperature
on the accretion parameters, and using the result that the \el~ luminosity is positively 
correlated with the radiation temperature, we can find the relation between the equivalent width
and the continuum luminosity for each pair of accretion scenario and temperature model.

\subsection {The L-T relation}

The dependence of the effective temperature on the \bh~ mass is  summarized in table 2. 
The header shows the three accretion scenarios and the 
corresponding dependence of the continuum luminosity on the mass.
The next row gives the dependence of the luminosity on the mass for each scenario.
The left column shows the temperature models. For each pair the table gives 
 the functional dependence of $T_{eff}$ on M.
Applying the dependence of L on M it is possible to deduce the dependence of T on L.

Combining the \bb -radiation boundary  and the photospheric models, which have a very 
similar dependence on the accretion parameters, the table has nine ($T_{eff},\dot M$) pairs,
for \ad models, and an additional row for the model independent \bb - variability 
estimate of $T_{eff}$ (marked ``$T$(BB-var)''). 

For all pairs except two  (constant $\dot M$ with $T({\rm BB-rad})$ and homogeneous medium 
accretion with $T_{BB}$).  The effective disk temperature is decreasing with increasing \bh~ mass.
For all accretion scenarios  the optical and UV continuum luminosity is increasing with $M$. 
This is certainly the case for \ed bounded accretion and for accretion from a homogeneous medium;
For a constant \ar~ the bolometric luminosity is constant, but as the \bh~ mass
increases the peak of the \ad spectrum moves to lower frequencies. Since the optical or 
near UV continuum are on the Rayleigh-Jeans part of the \bb spectrum,
the luminosity in these bands increases as the mass increases even if 
the total luminosity remains constant.

We note that this is true also for the general temperature-luminosity relation
 (eqs. \ref{equ:elt} and \ref{equ:el}), so  the \ad model is not essential.
In that case the anti-correlation between temperature and luminosity is built into the
the expression for the temperature, eq. \ref{equ:el}.

{\bf Decreasing \ar .}
In the cases we looked at the \ar~ increases with time or remains constant.
What if the accretion rate decreases with time? 
In the two latter \ad -temperature models (5.2.c and d) we find
$E_{max}\sim \dot M^{-1/3}$, so $T_{eff}$ increases, and so does the line luminosity.
If we take $L\propto \dot M$ the continuum luminosity decreases, and 
there is still a negative correlation.
This is the case also in the general L-T relation. For the first two models (5.2.a,b) the
situation is less clear, because $T_{eff}$ in these two models will decrease 
(as $\dot M^{1/4}/M^{1/2}$
and $M^{-1/4}$, respectively) and also $L\sim \dot M$ decreases. The question whether 
$T_{eff}$ is correlated or anti correlated with $L$ would depend on the details of the
behavior of the \ar .

\subsection {The \ew }
How does the \ew vary? The \ew is defined as $L_{line}/F_\lambda$,
where $F_\lambda$ is the flux per unit wavelength at the wavelength of the line.
We have argued that $L_{line}$ depends on the ionizing continuum, which correlates 
strongly and steeply with $T_{eff}$. The luminosity in the line is also roughly 
linearly dependent on the continuum luminosity, $L\sim \lambda F_\lambda$, so that
$$L_{line}\propto T_{eff} L ,$$
which gives for the \ew 
$$EW(L)\sim L_{line}/L\sim  T(M).$$

The last approximation should not be taken as a linear relation, but rather as
a strong correlation. As shown above, for $3kT<E_{ion}\sim$ few Ryd, $L_{ion}$ 
is steeply increasing (and hence strongly correlated) with $T$, and hence
$L$(line) is strongly correlated with the temperature.

\subsection {The Exceptions}
We conclude that for many accretion-scenarios and temperature-models
(accretion-disk models or general effective \bb related to variability)   
the line luminosity decreases with increasing mass, except
the two for which the effective temperature is constant, there is an evolutionary
\be : as the \bh~ evolves over cosmological times due to the mass accreted, the continuum
luminosity (optical or near UV) increases, while the \ew decreases.
Even the two pairs with a constant effective temperature will eventually show a \be .
To see this, note that for the ($T_{BB}-\dot M$(spherical)) pair, $\dot m$ is increasing
as $M$, so it will eventually approach the \ed \ar , and move from the $T_{BB}$ model
to one of the other $T_{eff}$ models,  which (for $\dot M$(spherical)) do have a \be .
In the other pair, (constant $\dot M$ with $T({\rm BB-rad})$), the situation is opposite,
but with the same result: since there $\dot m\sim M^{-1}$, eventually the \ar~ will become
enough sun-\ed so that the $T_{BB}$ model will apply, which for the constant \ar~
scenario does have a \be .

\section{Summary}
We suggest that
the growth of the central \bh~ mass due to the  accreted matter
causes  the luminosity and the spectrum to
evolve over over the AGN life time.
Using  general model-independent relations, or the thin \ad  spectrum,
we estimate the evolution of the \ew and the continuum spectrum, and
show that for plausible evolutionary tracks 
as well as for the model-independent \bb temperature estimate and for
most variants of the thin \ad model
the \ew decreases with increasing continuum luminosity
 implying an evolutionary origin to the Baldwin effect.

\end{document}